\documentstyle[12pt]{article}
\newcommand{\bra}[1]{\raisebox{-0.3ex}{\mbox{\tt $\langle$}}#1
\hspace{0.5ex}\rule[-0.5ex]{0.1ex}{0.9em}\hspace{0.5ex}}
\newcommand{\ket}[1]{\hspace{0.5ex}\rule[-0.5ex]{0.1ex}{0.9em}
\hspace{0.5ex}#1\raisebox{-0.3ex}{\mbox{\tt $\rangle$}}}
\renewcommand{\d}{\displaystyle}
\newcommand{\Kr}[1]{\left( #1\right)}
\newcommand{\lapprox}{\mbox{\raisebox{-4pt}{$\,\buildrel<\over\sim\,$}}}
\newcommand{\gapprox}{\mbox{\raisebox{-4pt}{$\,\buildrel>\over\sim\,$}}}
\newcommand{\ve}{\varepsilon}
\newcommand{\ds}[1]{_{\mbox{\scriptsize\rm #1}}}
\newcommand{\us}[1]{^{\mbox{\scriptsize\rm #1}}}
\renewcommand{\textfraction}{0}
\begin{document}
\setcounter{totalnumber}{99}
\renewcommand{\textfraction}{0}
\title{Wigner Molecules in Nanostructures}
\author{K.~Jauregui, W.~H\"ausler and B.~Kramer\\
Physikalisch--Technische Bundesanstalt Braunschweig\\
Bundesallee 100, 3300 Braunschweig F.~R.~G.}
\maketitle
\date{}
\begin{abstract}
The one-- and two-- particle densities of up to four interacting electrons
with spin, confined within a quasi one--dimensional ``quantum dot'' are
calculated by numerical diagonalization. The transition from a dense
homogeneous charge distribution to a dilute localized Wigner--type electron
arrangement is investigated. The influence of the long range part of the
Coulomb interaction is studied. When the interaction is exponentially cut off
the ``crystallized'' Wigner molecule is destroyed in favor of an inhomogeneous
charge distribution similar to a charge density wave .
\end{abstract}
\hspace{1cm}

\pagebreak

Single charges dominate the electronic properties of submicron structures at
low temperatures. Due to small capacitances the charging energy associated
with the addition of one electron into a given structure can exceed the
thermal energy. Two recently discovered important phenomena can be explained
by single electron charging effects: the Coulomb blockade of the dc--current
through small tunnel junctions \cite{averinschon,grabert}, and the periodic
oscillations of the conductance of quantum dots \cite{meirav,kastner}.

In contrast to metallic systems semiconductor nano--structures allow to
reduce the number of electrons in a quantum dot by changing a gate voltage.
Optical absorption experiments were performed on quantum dots which contain
only $\:N=2\ldots 4\:$ electrons \cite{heitmann}. For very low electron
densities additional effects in the transport properties can be expected due
to the increasing importance of the Coulomb interaction \cite{corr91} since
the electrons tend to ``crystallize'' into an inhomogeneous ground state
\cite{wigner}. Besides Coulomb effects, linear \cite{leothesis} and non--
linear \cite{johnson} transport experiments show fine structure in the
current voltage characteristics that can be traced back to the granularity of
the charge density distribution and to the lowest collective excitations
\cite{master}, respectively.

In this paper, we demonstrate that for a quasi--one  dimensional (1D) system
containing a few electrons the charge density $\varrho(x)$ becomes
inhomogeneous for sufficiently large mean electron distance $r\ds{s}$.
Our results indicate that there are two consecutive regimes of charge
localization. First, $\varrho(x)$ starts to become inhomogeneous with
increasing $r\ds{s}$. The corresponding energetically lowest excitations are
associated with changes in symmetry and total spin \cite{corr92}. When
$r\ds{s}$ is further increased, $\varrho(x)$ vanishes almost completely
between well separated maxima. In analogy with the Wigner lattice, we denote
this limiting configuration in a finite system as a ``Wigner molecule''. Two
classes of elementary excitations can be identified in this limit: tunneling
between different permutational arrangements of the electrons and vibrations
of the charge density \cite{corr91,corr92}.

We consider $N$ interacting electrons including spin, confined in a 1D square
well potential of finite length $L$. The height of the potential was assumed
to be finite but large. The Hartree--Fock approximation is known
\cite{mattis} to overestimate the ferromagnetic state and cannot reproduce
the true ground state which is antiferromagnetic in 1D \cite{liebmattis}.
We calculate eigenvalues and eigenstates of the Hamiltonian
exactly.

The kinetic energy
\begin{equation}
E\ds{H}\Kr{\frac{a\ds{B}}{L}}^2
\sum_{n,\sigma}\epsilon_nc^{\dagger}_{n,\sigma}
c^{\phantom{\dagger}}_{n,\sigma} \equiv E_{\rm H}\Kr{\frac{a_{\rm B}}{L}}^{2}
H_{0}
\end{equation}
and the Coulomb interaction
\begin{equation} \label{coul}
E\ds{H}\Kr{\frac{a\ds{B}}{L}}\sum_{n_{1}...n_{4}\atop\sigma,\sigma'}
V_{n_{4}n_{3}n_{2}n_{1}} c_{n_{4}\sigma}^{\dagger}
c_{n_{3}\sigma'}^{\dagger}c_{n_{2}\sigma'}c_{n_{1}\sigma}\equiv
E\ds{H}\Kr{\frac{a\ds{B}}{L}}H\ds{I}
\end{equation}
scale like $\:1/L^2\:$ and $\:1/L\:$, respectively. Thus the latter dominates
at large $L$. The natural energy and length scales are given by the Hartree
$\:E\ds{H}=e^2/a\ds{B}\:$,  and the Bohr radius,
$\:a\ds{B}=\ve\hbar^2/me^2\:$ ($\:\ve\:$ relative dielectric constant,
$\:m\:$ effective electron mass). $\:c^{\dagger}_{n,\sigma}\:$ creates a
particle with energy $\:\epsilon_n\:$ ($=\:(n\pi)^{2}/2\:$ for an infinitely
high well) and spin $\:\sigma\:$. $V_{n_{4}n_{3}n_{2}n_{1}}$ is the matrix
element of the interaction potential $V(x,x')={e^2}/{\ve \sqrt{(x-
x')^2+\lambda^2}}$. The cutoff at short distances simulates a small
transversal spread $\:\lambda\ll L\:$ of the wave functions and leaves the
$V_{n_{4}n_{3}n_{2}n_{1}}$ finite. In the calculations we assumed $\:{\lambda
/L=2\cdot 10^{-4}}\:$. The eigenvalues of
$H=E\ds{H}(a\ds{B}/L)[(a\ds{B}/L)H_0 + H\ds{I}]$ depend only weakly on
$\lambda$ when $\lambda\ll1$ \cite{corr91}. To investigate the effect of the
long range part of the interaction, in some of the calculations an
exponential cutoff at large distances was introduced. No charge neutralizing
background was taken into account since it is not expected to modify the
results for fixed $\:N\:$ qualitatively.

The Hamiltonian matrix in the basis of the Slater determinants was
diagonalized numerically. The latter were constructed from the single
electron states $\:\varphi_{n}(x)\chi_{\sigma}\:$ where $\varphi_{n}$
is a spatial function and $\chi_{\sigma}$ a spinor with
$\:\sigma=\downarrow,\uparrow\:$. Considering $1\le n\le M$ leads to a matrix
of the rank $\:\d R=\rule[-3ex]{0ex}{7.2ex} \Kr{2M\atop N}\:$ ($R<1.5\cdot
10^4$, for $\:M=10\ldots 17\:$). We used Lanczos procedures when $\:R>3\cdot
10^{3}\:$.

The one-- and two--particle densities were determined from the normalized
eigenfunctions $\psi (x_{1},...,x_{N}) = \sum_{\nu=1}^{R}
\:({\mbox{b}_{\nu}}/{\sqrt{N!}})\;\mbox{det}\{\varphi_{n_{i}}(x_{j})
\chi_{\sigma_{i}}\}$, were $\:\nu\:$ denotes the configurations
$\:\{n_{1}\sigma_{1},...,n_{N}\sigma_{N}\}\:$. The coefficients
$\mbox{b}_{\nu}$ can be chosen real. In second quantization the one--particle
density is given by
\begin{equation}\label{dens}
\varrho (x) =\sum_{\sigma}\bra{\psi^{(N)}_{0}}\Psi_{\sigma}^{\dagger}(x)
\:\Psi_{\sigma}^{\phantom{\dagger}}(x)\ket{\psi^{(N)}_{0}} \quad ,
\end{equation}
were $\:\ket{\psi_{0}^{(N)}}=\sum_{\nu=1}^{R} \mbox{b}_{\nu}^{0}\ket{\nu}\:$
is the $\:N$--electron ground state. $\Psi_{\sigma}^{\dagger}(x)=
\sum_{n=1}^{M}\varphi_{n}(x)c_{n,\sigma}^{\dagger}\:$ creates an electron at
position $x$ with spin $\sigma$ and
$\:\ket{\nu}=c_{n_{1}\sigma_{1}}^{\dagger}\cdots
c_{n_{N}\sigma_{N}}^{\dagger}\ket{0}\:$ is a noninteracting Slater
determinant. Straightforwardly we have

\begin{equation}
\varrho (x) = \sum_ {\nu =1 \atop \nu'=1}^{R}
\mbox{b}_{\nu}^{0}\mbox{b}_{\nu'}^{0} \sum_{n,n' \atop \sigma} \varphi_{n}(x)
\varphi_{n'}(x) \bra{\nu'}c_{n',\sigma}^{\dagger} c_{n,\sigma} \ket{\nu}
\quad .
\end{equation}

The development of maxima and minima in $\varrho(x)$ with increasing mean
electron distance $r_{\rm s}\equiv L/(N-1)$ is shown in figure 1 for
$\:N=3,4\:$. The finite height of the square well potential was assumed to be
proportional to $1/L^{2}$ such that the number of non--interacting bound
states was independent of $L$. This implies a non--vanishing but small
$\varrho(x)$ outside the box.

We distinguish three characteristic regimes of electron densities. For small
distances, $\:r_{\rm s}\lapprox 0.1a\ds{B}\:$, the Coulomb interaction
perturbs $\:E\ds{H}(a\ds{B}/L)^{2}H_0\:$ only weakly such that $\varrho(x)$
is dominated by the lowest occupied single particle states. This explains
the minimum at $\:x=0\:$ (Fig.1). $\varrho(x)\:$ changes qualitatively at
$\:r_{\rm s}\approx a\ds{B}$. A structure consisting of $\:N$ peaks emerges.
The ``critical'' length for this transition is of the same order as found in
\cite{corr92} for the transformation from an almost non--interacting energy
spectrum to the multiplet spectrum dominated by interaction. When $r\ds{s}$
increases further, say $\:r_{\rm s}\gapprox 100a\ds{B}\:$, $\varrho(x)$
vanishes almost completely in finite regions between the maxima indicating a
fully established Wigner molecule. In this limit the ground state energy can
be approximated by that of a chain of elementary charges at equal distances
$\:r_{\rm s}\:$.

In order to investigate the influence of the long range part of the Coulomb
interaction, we introduced an exponential cutoff $V(x,x')\propto{\mbox{e}^{-
\alpha |x-x'|}}/{\sqrt{(x-x')^2 +\lambda^2}}$, were $\:1/\alpha\:$ is the
cutoff distance. The inset of figure 1 shows $\:\varrho(x)\:$ for $\:N=3\:$
and $\:\alpha =10a\ds{B}^{-1}\:$. Despite the extremely short range of the
interaction pronounced maxima are observed. However, in contrast to $\:\alpha
=0\:$ the density between the maxima is much higher even for very large
$r\ds{s}$. The inset of figure 1 rather resembles a charge density wave
\cite{cdw}. We conclude that the long range part of the Coulomb interaction
is essential for the charge condensation and for long range density--density
correlations. The ground state energy for $\:\alpha > r_{\rm s}^{-1}\:$ does
not converge towards the Coulomb energy of $N$ elementary point charges for
$\:L\rightarrow\infty\:$.

For large $\:r\ds{s}$ $(r\ds{s}\gg a\ds{B})\:$ the spectrum consists of
multiplets \cite{multipl}. Their energetic separation $\Omega$ correspond to
vibrational excitations. Their internal splitting $\Delta$ is due
to tunnelling through the Coulomb barrier between adjacent permutational
arrangements of the electrons \cite{corr92}. Asymptotically, $\:\Omega\:$
decreases as $r_{\rm s}^{-\gamma}$. Power law behavior is also obtained
for $\:\alpha>0\:$. However, the exponent $\:\gamma\:$ changes from
$\:\gamma\lapprox 1.5\:$ ($\alpha=0$) to $\:\gamma\approx 2\:$ ($\alpha>0$)
(figure~2). The vibrational excitations are independent of the symmetry of
the spatial part of the $N$--electron wave function and of the total spin. In
analogy to the argument for $\alpha=0$ leading to $1\le\gamma\le 1.5$
\cite{corr92}, the almost flat two--particle potential between adjacent
charge density peaks for $\:\alpha\gg r_{\rm s}^{-1}\:$ leads to
$\:\gamma=2\:$. Each electron is practically moving freely within a box of
length $r\ds{s}$. On the other hand, $\Delta$ decreases roughly exponentially
with increasing $L$, even for $\:\alpha>0\:$ (inset of figure~2). However,
$\Delta$ is larger and the decrease is less rapid when $r\ds{s}$ is increased
than for $\alpha=0$. The enhancement results from the reduced thickness of
the screened Coulomb barrier such that the corresponding tunnelling integral
($\propto\Delta$) increases. The modified dependence on $r\ds{s}$ is due to
the fact that only for $\:r_{\rm s}<\alpha^{-1}\:$ the thickness of the
interaction barrier varies with $\:r_{\rm s}\:$. We find that the tunnelling
excitations can be described in terms of correlated, localized basis
functions even for $\:\alpha>0\:$. However, in order to obtain similar
quantitative agreement with the numerically determined eigenvalues as for
$\:\alpha=0\:$ a larger mean electron distance $\:r_{\rm s}\:$ is required to
fulfill $\:\Delta\ll\Omega\:$. As a consequence the characteristic length for
the transition from almost non--interacting to the interaction dominated
behavior increases with increasing $\:\alpha\:$.

The pronounced correlations between the electrons in our model are also
observed in the pair correlation function of the $\:N$--electron state
$\:\ket{\psi^{(N)}}\:$,
\begin{equation}\label{twodef}
\varrho\ds{c}(x,x') = \sum_{\sigma,\sigma'}
\bra{\psi^{(N)}}\Psi_{\sigma}^{\dagger}(x)\: \Psi_{\sigma'}^{\dagger}(x')
\:\Psi_{\sigma'}^{\phantom{\dagger}}(x')
\:\Psi_{\sigma}^{\phantom{\dagger}}(x)\ket{\psi^{(N)}} \quad .
\end{equation}
$\varrho\ds{c}$ is related to the density--density correlation function
$\:\varrho_2\:$ via $\:\varrho_2(x,x') = \varrho(x)\delta(x-x') +
\varrho\ds{c}(x,x')\:$. Physically, $\:\varrho\ds{c}\:$ is the probability to
find an electron at position $\:x\:$ when another electron sits at $\:x'\:$
($\:\int\mbox{d}x\int\mbox{d}x'\:\varrho\ds{c} (x,x')=N(N-1)\:$). Figure 3
shows a contour plot of $\:\varrho\ds{c}\:$ for the $\:N=3\:$ ground state
($S=1/2$) with $L=9.5a_{\rm B}\:$. Practically the same result is
obtained for the spin polarized excited state ($S=3/2$). The difference
between $\varrho\ds{c}$ of the ground state and the spin polarized state is
less then 1\%, independent of $x$, and $x'$.

It is known that the Hartree Fock approximation overestimates the tendency of
electrons with parallel spins to avoid each other and underestimates the
correlations between electrons with antiparallel spins. Therefore large
differences between correlation functions for different spin configurations
are expected in this approximation. Recently Pfannkuche et al. demonstrated
this property explicitly for three electrons in a two dimensional parabolic
potential \cite{daniela}. The weak sensitivity of the pair correlation
function on $S$ shows that highly correlated electrons in a ``slim'' quantum
dot cannot be treated within molecular field approximation.

In summary, we have analyzed one-- and two--particle densities for the lowest
eigenstates of $\:N=2,3,4\:$ electrons confined in a quasi--1D square well.
Three different regimes were identified. At high electron densities the
Coulomb interaction is only a weak perturbation. Between $\:a\ds{B}^{-
1}\gapprox r_{\rm s}^{-1}\gapprox 10^{-2}a\ds{B}^{-1}\:$ the electrons tend
to localize. Simultaneously the eigenvalue spectrum develops a multiplet
structure that is dominated by the interaction. Both, the inhomogeneity of the
charge density and the multiplet character of the spectrum are not destroyed
when the long range part of the interaction is removed. However, there are
quantitative changes in the asymptotic behavior of tunneling and vibrational
excitations at low densities. For the lowest densities (equivalent to
$r\ds{s} \gapprox 100a\ds{B}$) the formation of a {\em Wigner molecule\/} is
observed. In this limit where the long range part of the interaction is
crucial, $\varrho(x)$ vanishes within finite intervals between the maxima.
The structure in the density--density correlation function becomes more
pronounced with increasing influence of the interaction but is less sensitive
to total electron spin than expected within Hartree Fock approximation. It is
worth remarking that the best numerical estimate to our knowledge for the
critical value of $r\ds{s}$ for the formation of a Wigner cristal in 3D is
also $r\ds{s}\us{cr}\approx 100a\ds{B}$ \cite{ceperley}.

Experiments are frequently performed on 2D quantum dots that are based on
GaAs--AlGaAs heterostructures. In many of these the electron density
corresponds to the intermediate regime. For the geometry and the electron
numbers in typical quantum dots \cite{meirav,kastner} $\:r_{\rm s}\approx
3a\ds{B}\:$ (area of the dot $\:\approx 10^5\:$nm$^2\:$, number of electrons
$\:\approx 10^2\:$, effective mass $\:\approx 0.07m\ds{e}\:$, dielectric
constant $\:\approx 10$). Even for this relatively high electron density in
comparison with more recent experiments \cite{heitmann} the charge density
distribution cannot be expected to be homogeneous if we assume that the
qualitative features of the electronic properties discussed above apply also
to the 2D case. In small quantum dots that are weakly coupled to leads and
that contain only a finite number of electrons, like the experimentally
investigated double barrier systems \cite{meirav}, a granular charge density
should lead to characteristic features in linear transport properties as a
function of the gate voltage $\:V\ds{G}\:$. A variation of $\:V\ds{G}\:$ not
only changes the charge density or the electron number inside the dot but
probably simultaneously its shape. In \cite{kastner} the gate electrode is
positioned besides the dot region. For this geometry the influence of
$\:V\ds{G}\:$ on the shape of the dot area is obvious. The resonance
condition for the appearance of a conductance peak in linear transport
$E\ds{F}=E_{N+1}-E_N$ may be fulfilled more than once for given $\:N\:$ when
$\:V\ds{G}\:$ is increased. Here $\:E\ds{F}\:$ is the chemical potential of the
leads (where no electron--electron interaction is assumed) and $\:E_N\:$ the
$\:N$--electron {\em ground state} energy of the quasi--isolated dot. Such a
model could account for the experimentally observed fine structure in the
conductance peaks in the {\em linear} transport regime reported in
\cite{leothesis} which seem to be presently not yet very well understood.

This work was partly supported by grants of the Deutsche
Forschungsgemeinschaft (AP 47/1-1) and EEC (Science Contract No.SCC$^{*}$--
CT90--0020).

\newpage
\begin{figure}[h]
\parbox[b]{15cm}
{\caption[onepart]{\label{onepart}
(a) One--particle density $\:\varrho (x)\:$ for $N=3$ and (b) $N=4$ for
different $L$ ($\:0.1a\ds{B}\le L\le 945a\ds{B}\:$, $M=13$). The
normalisation is such that $\:\int\mbox{d}x\:\varrho (x)=N\:$. When
$\:L\gapprox 1a\ds{B}\:$ $N$ peaks begin to emerge. For $\:L\gapprox
100a\ds{B}\:$ the peaks are well separated. Inset: $\varrho(x)$ for a pair
potential with an exponential cutoff ($\:\alpha=10a\ds{B}^{-1}\:$) for $N=3$
and $L=0.1a\ds{B}$ (dotted line); $L=9.5a\ds{B}$ (dashed--dotted line);
$L=472a\ds{B}$ (dashed line); $L=1417a\ds{B}$ (solid line) ($\:M=11\:$). The
minima at large system lengths are less pronounced as for $\alpha=0$.

}}
\end{figure}
\begin{figure}[h]
\parbox[b]{15cm}
{\caption[figcutoff]{\label{figcutoff}
Influence of a pair potential on the energy difference between the lowest two
multiplets $\:\Omega\:$ (vibrational excitation). The ratio
$\:\Omega(\alpha=10a\ds{B}^{-1})/\Omega(\alpha=0)\:$ is plotted versus the
particle distance $r_{\rm s}$ for $\:N=2\:$ ($\times$) and $\:N=3\:$
($\circ$). ($M=11$). Inset: logarithm of the energy difference $\:\Delta\:$
between the ground state and the first excited state within the lowest
multiplet versus system length $\:L\:$ for $N=2$, $\alpha=0$ ($+$) and
$\alpha=10a\ds{B}^{-1}$ ($\circ$); $N=3$, $\alpha=0$ ($\times$) and
$\alpha=10a\ds{B}^{-1}$ (Y). ($M=11$).

}}
\end{figure}
\begin{figure}[h]
\parbox[b]{15cm}
{\caption[twopart]{\label{twopart}
Contour plot of the pair correlation function $\:\varrho\ds{c}(x,x')\:$
(cf eq.~\ref{twodef}) for the $\:N=3\:$ ground state ($\:S=1/2\:$) at
$\:L=9.5a\ds{B}\:$ ($M=13$).}}

\end{figure}
\newpage

\newcommand{\pap}[5]{#1,\ #2{\bf #3}, #4 (19#5)}


\begin{thebibliography}{99}
\bibitem{multipl}There are $\:2^N\:$ (equal to the dimension of the
  spin Hilbert space) states per multiplet, differing in total spin and in
  symmetry with respect to permutations of the spatial space coordinates of the
  $\:N$--electron wave function \cite{corr92}.
\end{thebibliography}
\end{document}